\documentclass[aps,pre,reprint,groupedaddress,showpacs]{revtex4-1}

\usepackage{graphicx}
\usepackage{dcolumn}
\usepackage{bm}

\usepackage{amsmath}
\usepackage{amssymb}
\usepackage{balance}
\usepackage{color}
\usepackage{lastpage}
\usepackage{soul}

\begin{document}

\preprint{APS/xxxx}

\newcommand{\cs}[1]{{\color{red}#1}}
\newcommand{\rd}[1]{{\color{blue}#1}}
\newcommand{\sh}[1]{{\color{green}#1}}

\title{Filling transitions on rough surfaces: inadequacy of Gaussian surface models}

\author{Renaud Dufour}
\author{Ciro Semprebon}%
\author{Stephan Herminghaus}
 \email{stephan.herminghaus@ds.mpg.de}
\affiliation{%
 Max Planck Institute for Dynamics and Self-Organisation, 37077 G\"{o}ttingen, Germany}%

\begin{abstract}
We present numerical studies of wetting on various topographic substrates, including random 
topographies. We find good agreement with recent predictions based on an analytical 
interface-displacement-type theory \cite{Herminghaus2012, Herminghaus2012a}, except that we find critical end points  
within the physical parameter range. As predicted,  Gaussian random surfaces are found to behave qualitatively different from 
non-Gaussian topographies. This shows that Gaussian random processes as models for rough surfaces must be used with great care, if at all,  
in the context of wetting phenomena.
\end{abstract}

\pacs{68.05.-n, 05.40.-a, 64.75.-g, 68.08.-}

\maketitle

\section{\label{sec:intro}Introduction}

While the physics of wetting and spreading on ideally smooth surfaces has reached a status 
of mature textbook knowledge, the wetting properties of randomly rough solid substrates, 
which are by far more relevant, are still poorly understood.
This is in part due to the wide range of length scales involved. Its lower bound is the 
characteristic length scale of the molecular interactions shaping the 
three-phase contact line region, which is on the order of a few tens of nanometers. 
Topographic roughness or chemical inhomogeneities on length scales smaller than that 
are integrated by these forces into quantities like the contact angle 
or the contact line friction. Its upper bound is set by the capillary length, 
above which gravity effects become dominant. It is usually of the order of a millimeter 
(for water, it is about $2.7$ mm). The lateral scale of typical roughness topographies of 
natural surfaces is quite generally found within this range, which extends over five 
orders of magnitude. 
 
Despite its obvious importance, this range of random topographies has so far only scarcely 
been treated as to its effect on wetting.  Most authors have tried to focus on isolated 
aspects of the topography using simplified model geometries, like rectangular \cite{Seemann2005}, 
triangular \cite{Rejmer1999}, or algebraic grooves \cite{rascon2000}. It has been found that at a certain contact angle and vapor pressure, a filling transition 
occurs at which the formerly dry grooves are filled with liquid \cite{parry2013, parry2014, parry2015}. 
The transition can be continuous or discontinuous, depending
on the geometry and the surface energy of the (regular) topography.
 
Whenever the random nature of the roughness was taken into account, 
surfaces were usually modeled by Gaussian random processes 
\cite{Persson2005, Grigorios2010, Chakraborty2012, Wilhelm2013, Carbone2015}, 
but filling transitions on randomly rough surfaces have been addressed only very recently.  
On the basis of an interface displacement model, it has been suggested  
that filling transitions are generic for a wide range of roughness 
occurring naturally \cite{Herminghaus2012, Herminghaus2012a}. 
Corresponding phase diagrams have been given, based on certain approximate assumptions.
Remarkably, it was predicted that it is precisely the Gaussian surfaces which behave qualitatively 
different from all other random topographies. Since no real surface is precisely Gaussian, 
this has potentially wide ranging consequences. 
 
Given the practical importance of this prediction, deeper  investigations are necessary to test the predictive power of   
the interface displacement theory, and thereby to verify its inherent assumptions.
In the present paper, we investigate the wetting phase diagrams for a number of different 
surface topographies, including simple model geometries as well as random surface profiles of real samples. 
In accordance with the predictions mentioned above, we find that it is 
essentially the non-Gaussian features which gives rise to filling transitions. 
 
The first systematic study of wetting on an irregularly 
rough substrate at finite contact angle is due to Wenzel \cite{Wenzel1936}. He characterized the 
roughness by a single parameter, $r$, which he defined as the ratio of the total substrate area 
divided by the projected area. The free energy which is gained per unit area when the rough 
substrate is covered with a liquid is given by $r(\gamma_{sg} - \gamma_{sl})$ (subscripts $sl$ 
and $lg$ standing for solid-liquid and solid-gas interfaces, respectively). If this is larger 
than the surface tension of the liquid, $\gamma$, the macroscopic contact angle vanishes, 
because covering the substrate with liquid releases more energy than is required for the 
formation of a free liquid surface of the same (projected) area. More specifically, force 
balance at the three-phase contact line yields
\begin{equation}
\label{eq.wenzel}
\cos\Theta=\frac{r(\gamma_{sg}-\gamma_{sl})}{\gamma}=r\cos\theta
\end{equation}
for the macroscopic contact angle, $\Theta$, on the rough surface. $\theta = \arccos\{(\gamma_{sg}-\gamma_{sl})/\gamma \}$ is the microscopic contact 
angle according to Young and Dupr\'{e}. When this is smaller than  
$\theta_W =  \arccos(1/r)$, which we will henceforth call Wenzel's angle, $\Theta$ 
vanishes, and the substrate is likely to be covered with an `infinitely' thick liquid 
film. We focus here on the vicinity of this transition, considering liquid-vapor coexistence, 
but also the off-coexistence case (i.e. below the saturated vapor pressure).

The article is organized as follows. In section \ref{sec:model} we briefly recall the 
framework of the interfacial displacement model on which we base our analysis.
In section \ref{sec:mumerical_minimization} we introduce the numerical method to compute 
equilibrium morphologies of the liquid interface on irregular surfaces.
In section \ref{sec:egg_carton}, we consider a simple periodic surface 
topography and compare the numerical results to the results of the interface displacement theory.
In section \ref{sec:irregular} both methods are applied to 
more complex, irregular substrates in order to confirm the generic character of the filling  transition.
Finally, in section \ref{sec:random_roughness}, we apply the interfacial displacement 
model to investigate the filling transition of a real physical roughness, as well as one having nearly 
Gaussian properties.

\section{\label{sec:model}interface displacement model}

Let us briefly review the interface displacement theory which has been put forward 
before \cite{Herminghaus2012a}. The aim of the theory is  
to predict the amount of adsorbed liquid as function the microscopic contact angle and the 
mean curvature of the liquid interface.

\begin{figure}
  \centering
  \includegraphics{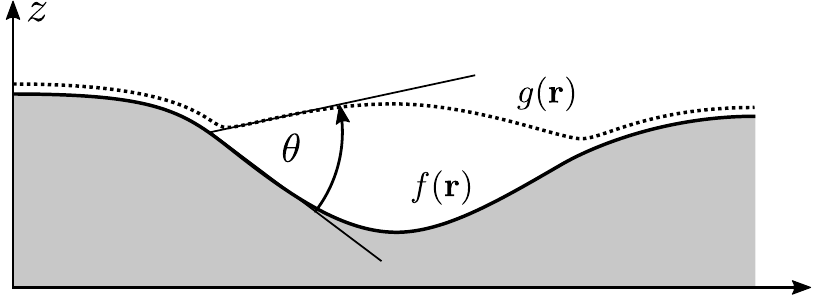}
  \caption{Substrate topography $g(\mathbf{r})$ and interface of
    the adsorbed liquid film $f(\mathbf{r})$ (dots) of (very small) finite thickness $l_0$.
    The microscopic contact angle is indicated by $\theta$.}
  \label{fig:surface_topography_and_film}
\end{figure}

We consider a homogeneous and isotropically rough solid surface defined on a set $\Omega$ 
in the plane by a function $z = f(\mathbf{r})$, where $z$ is the height of the surface 
above the point $\mathbf{r}=(x,y)$. We assume that $\nabla f$ and $\Delta f$ exist 
everywhere, and that the roughness amplitude is much smaller than its lateral length 
scale, $\nabla f \ll 1$, which is the case for a large class of  natural roughness. 
The free surface of a liquid film adsorbed on this substrate is described by a second 
function, $g(\mathbf{r})$, which deviates appreciably (i.e., by more than $l_0$) only in the wet area, $\mathcal{W}\in \Omega$. 
The morphology of the liquid/vapor interface, and thereby the amount of adsorbed liquid, 
is determined by the boundary conditions
\begin{eqnarray}
& g(\mathbf{r}) = f(\mathbf{r})+l_0 \quad & \forall \mathbf{r}\in \partial \mathcal{W} \nonumber \\
& |\nabla(g(\mathbf{r})-f(\mathbf{r}))| \approx \theta \quad 	&\forall \mathbf{r}\in \partial \mathcal{W} 
\label{eq.boundary_conditions}
\end{eqnarray}
where $g-f \gg l_0$ on the wetted patches of the sample. The last equation is to be fulfilled to first order in $\theta$, 
$\nabla f$ and $\nabla g$ everywhere on $\partial \mathcal{W}$, which is the projection 
of the three-phase contact line. Applying Green's theorem  to $(g-f)$, we can write
\begin{equation}
\int_{\partial \mathcal{W}} \mathbf{n}\cdot\nabla(g-f) \, \mathrm{d}s = \int_{\mathcal{W}} \Delta(g-f) \, \mathrm{d}^2\mathbf{r}
\label{eq.green}
\end{equation}
where $s$ is the distance along $\partial\mathcal{W}$ and $\mathbf{n}$ its unit normal vector. Writing Eq.~(\ref{eq.boundary_conditions}) as $\mathbf{n}\cdot\nabla(g-f) \approx \theta$ on $\partial\mathcal{W}$ and combining with Eq.~(\ref{eq.green}), we obtain
\begin{equation}
l\theta + \int_{\mathcal{W}} [2H - \Delta f] \, \mathrm{d}^2\mathbf{r} = 0
\label{eq.recast}
\end{equation}
where $l$ denotes the length of $\partial \mathcal{W}$ per unit surface area and $H$ is the mean curvature of the liquid/vapor interface.
In order to evaluate Eq.~(\ref{eq.recast}), it is worth describing a given topography in terms of statistical quantities. Therefore we define:
\begin{eqnarray}
p(z) 	&=	|\Omega|^{-1} \int_{\Omega} \delta(z-f(\mathbf{r})) \, \mathrm{d}^2\mathbf{r} 
\label{eq.pz}\\
\sigma_1(z)	 &=	p(z)^{-1} |\Omega|^{-1} \int_{\Omega} |\nabla f(\mathbf{r})| \, \delta(z-f(\mathbf{r})) \, \mathrm{d}^2\mathbf{r}
 \label{eq.sigma1} \\
\sigma_2(z)	 &=	p(z)^{-1} |\Omega|^{-1} \int_{\Omega} |\nabla f(\mathbf{r})|^2 \, \delta(z-f(\mathbf{r})) \, \mathrm{d}^2\mathbf{r}
 \label{eq.sigma2}
\end{eqnarray}
where $p(z)$ is the surface height distribution. $\sigma_1(z)$ and $\sigma_2(z)$ 
describe the average slope $\langle \vert  \nabla f\vert \rangle_z$ and average square slope 
$\langle \vert  \nabla f\vert^2 \rangle_z$, respectively, at elevation $z$. Equation (\ref{eq.recast}) 
can then be re-written as \cite{Herminghaus2012a}:
\begin{equation}
\label{eq.approximation}
2 \, H \, \int_{-\infty}^z p(f) \, \mathrm{d}f \approx \left[  \frac{\sigma_2(z)}{\sigma_1(z)} - \theta \right ] \, \sigma_1(z) p(z)
\end{equation}
We finally define the cumulative height distribution, $W(z) = \int_{-\infty}^z p(f)dz$ and note that the length of a contour line at elevation $h$ is given by   
$L(h) = \sigma_1(h) p(h)$\cite{Higgins1957}. It is one of the 
major approximations used previously \cite{Herminghaus2012,Herminghaus2012a} to assume that $L(h)$ is roughly equal to the length of the contact line if $h$ is taken to be the average height of that line, and that $W(h)$ is approximately equal to the wetted sample area.
We then obtain
\begin{equation}
\label{eq.master_equation}
2 \, H \, W(h) \approx \left[  \frac{\sigma_2(h)}{\sigma_1(h)} - \theta \right ] \, L(h).
\end{equation}
If $p(z)$, $\sigma_1(z)$ and $\sigma_2(z)$ are known from experimental characterization of the sample, Eq.~(\ref{eq.master_equation}) allows to determine 
the equilibrium level $h$ of the liquid filling the troughs, as a function of the microscopic angle 
$\theta$ and curvature $H$. A reasonable estimate of the volume of adsorbed liquid per unit area can then be obtained by
\begin{equation}
\label{eq.volume}
V(h) = \int_{-\infty}^{h} (h-z) \, p(z) \, \mathrm{d}z
\end{equation}
note that such approximation neglects the curvature 
of the liquid meniscus with respect to the horizontal plane defined by the cut at the level $z$.
Nevertheless, the behaviour of $V$ thus defined will qualitatively reflect all features of interest of an adsorption isotherm.

\section{\label{sec:mumerical_minimization}Numerical energy minimizations}

In order to assess the accuracy of the interface displacement model, we computed 
equilibrium profiles $g(\mathbf{r})$ of the free liquid interface 
from numerical minimizations of the free energy:
\begin{equation}
 \begin{aligned}
  {\cal F}\{g(\mathbf{r})\}=  &
  \frac{\gamma}{2}\,\int_\Omega {\rm d}^2r \;|\boldsymbol{\nabla} g(\mathbf{r})|^2 \\
   & + \int_\Omega{\rm d}^2r\;\Phi\left(\ell(\mathbf{r})\right)
  - \lambda \int_\Omega{\rm d}^2r\;\ell(\mathbf{r})~,
   \end{aligned}
  \label{eq:interfacial_free_energy}
\end{equation}
with the interfacial tension of the liquid-air interface, $\gamma$, the
local thickness of the liquid film, $\ell(\mathbf{r})\equiv g(\mathbf{r})-f(\mathbf{r})$, and the domain of
integration, $\Omega$. A sketch of a cross section
perpendicular to the substrate is shown in
Fig.~\ref{fig:surface_topography_and_film}. The parameter $\lambda$ in
front of the last term on the RHS of Eq.~(\ref{eq:interfacial_free_energy}) 
represents the Laplace pressure of a reservoir that can exchange
liquid with the substrate, i.e., $\lambda = 2\gamma H$.

To obtain equilibrium profiles with a desired apparent contact angle,
we use a generic short ranged effective interface
potential,
\begin{equation}
  \Phi(l)=\Phi_0\left(\frac{2\ell_0}{\ell(1+\sqrt{5})}-1\right)\exp\left(-\frac{\ell(1+\sqrt{5})}{2\ell_0}\right)
  \label{eq:interface_potential}
\end{equation}
The equilibrium film thickness $\ell_0$ corresponds to 
$\Phi^\prime(l_0)=0$ (vanishing disjoining pressure).
The depth of the global minimum of interface potential 
determines the apparent contact angle $\cos\theta=1-\mid \Phi(\ell_0)\mid /\gamma$.

For a given substrate topography $f(\mathbf{r})$, the free energy
(Eq.~(\ref{eq:interfacial_free_energy})) is numerically minimized
employing the public domain software Surface Evolver \cite{Brakke1996}. To this end, the free
interface is represented as a mesh of small triangles the nodes of which 
are moved with a conjugate gradient algorithm. The triangulation is 
regularly re-meshed, such as to keep the typical size of the edges below about $\sim \Lambda/20$, 
where $\Lambda$ is the shortest wavelength of the 
topography, and to avoid the formation of long and narrow triangles.
$\ell_0$ was set equal to $\Lambda/20$. It was found that the numerical stability could be further  
improved by introducing an additional hard repulsive constraint ensuring $\ell \ge \ell_0$. 

In a typical run, the computation is initialised by bringing the liquid interface
in touch with the highest peak of the roughness. The contact angle $\theta$
is maintained constant, while a small but finite pressure is applied.
By ramping up the pressure parameter $\lambda$ (Eq.~(\ref{eq:interfacial_free_energy})), 
we explore sequences of interfacial configurations of the liquid with a continuously 
varying mean curvature, $H$. The physical reason for this preference is that if one starts instead from 
a small amount of adsorbed liquid, coresponding to large interfacial curvature, the liquid consists of 
a large number of isolated liquid puddles, and it is not straightforward 
how to generate or select a well defined initial state. 
 
Note that the sequences of morphologies
obtained with this procedure do not necessarily represent global energy minima. This is resembles an experiment where the partial pressure of the vapour of the adsorbed liquid is gradually reduced. Given the fact that the typical length scales of natural roughness correspond to interfacial energy barriers much larger than thermal energies, global free energy minima are reached in experiments only after close to infinite times. It should be noted that a similar approach was already successfully adopted to investigate the instabilities of a liquid meniscus leading
to advancement of the liquid front on a regularly patterned substrate \cite{Semprebon2012}.

\begin{figure}
\includegraphics{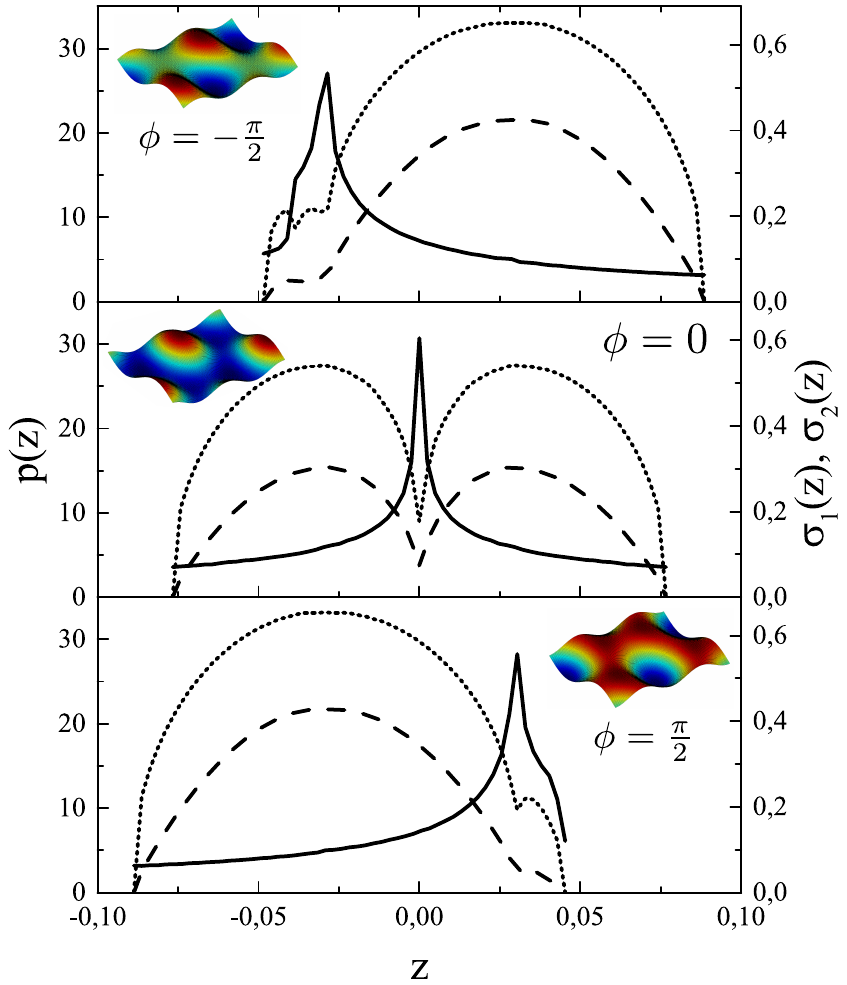}%
\caption{Statistical properties of the regular topography from Eq.~(\ref{eq.egg.carton}). From top to bottom: 
$\phi=(-\pi/2,\phi=0,\pi/2)$. Solid curves represent the height distribution 
$p(z)$. Conditional mean slope $\sigma_1(z)$ and $\sigma_2(z)$ are shown as 
dotted and dashed lines, respectively. Inset represents the unit cell of the periodic surface.}
\label{fig.stats_egg_carton}
\end{figure}

\section{\label{sec:egg_carton}egg-carton topography} 

The major difference between one dimensional periodic topographies and
truly two dimensional substrates is the presence of saddle points
in between peaks and valleys in the latter. To assess the fundamental mechanisms 
underlying the drying and filling transition on two-dimensional 
topographies, we begin our analysis considering a simple example. It consists of a 
hexagonal arrangement of peaks, valleys and saddle 
points, whose locations can be set by tuning appropriate parameters. 
We will refer in the following to it as `egg carton'. 
An hexagonal egg-carton can be mathematically described by
\begin{equation}
z = A\left( \sin(\frac{4\pi u_1}{L}+\phi)  +\sin(\frac{4\pi u_2}{L})+\sin(\frac{4\pi u_3}{L}) \right)
\label{eq.egg.carton}
\end{equation}
where $L$ is the unit length of the periodic cell, $A~=~0.03\times~L$ tunes the amplitude of the corrugation and $u_i$ represent hexagonal coordinates,
\begin{eqnarray}
u_1 &=  &x						\nonumber \\
u_2 &=  &\frac{x + \sqrt{3}y}{2}	\nonumber \\
u_3 &= -&\frac{x - \sqrt{3}y}{2}
\end{eqnarray}
In Eq.~(\ref{eq.egg.carton}), the phase variable $\phi$ enables to tune the vertical level of the saddle point.
We consider three different configurations, with $\phi = -\pi/2$, $\phi = 0$, and $\phi =\pi/2$. 
These choices place the saddle point close to the top, in the mid-plane,  and close 
to the bottom of the topography, respectively. It can be observed in Fig.~\ref{fig.stats_egg_carton} 
that the level of an isolated saddle point significantly affects the shape of the distributions 
$p(z)$, $\sigma_1(z)$ and $\sigma_2(z)$ (cf. Eq.~(\ref{eq.pz}), (\ref{eq.sigma1}), and (\ref{eq.sigma2}), respectively).

\begin{figure}[tb]
\includegraphics{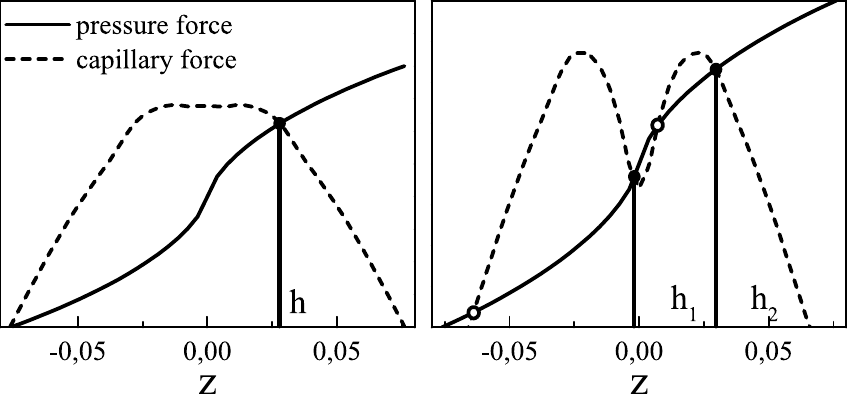}
\caption{Graphical solution of Eq.~(\ref{eq.master_equation})
for the egg-carton pattern with $\phi=0$ (Fig.~\ref{fig.stats_egg_carton} middle panel) and 
two combinations of $[\theta, H]$ values. Left : $\theta=5^{\circ}$ and $H=1$.
This configuration admits a single solution at $z\approx 0.025$. Right : $\theta=20^{\circ}$ 
and $H=0.4$. This configuration admits multiple solutions of which two are stable (closed circles) 
and two are unstable (open circle). The stability of the solution depends upon the slopes of the intersecting curves.}
\label{fig.graphi_solution_egg} 
\end{figure}

The average level $h$ of the contact line of an adsorbed liquid film can be 
predicted with the interface displacement approach by solving Eq.~(\ref{eq.master_equation}).
For given values of $\theta$ and the mean curvature of the free liquid surface, $H$, 
the solution can be obtained graphically. To illustrate the mechanism employed 
in the numerical solution of Eq.~(\ref{eq.master_equation}), let us consider the case 
$\phi=0$ in the examples depicted in Fig.~\ref{fig.graphi_solution_egg}. Depending 
on $\theta$ and $H$, Eq.~(\ref{eq.master_equation}) admits either single or multiple 
solutions. The stability of the solutions given by the graphical intersection can 
be inferred by considering that Eq.~(\ref{eq.master_equation}) represents a force balance. 
For a solution to be stable, upon a displacement of the three-phase contact line, the unbalance 
of wetting forces have to bring the film back toward the equilibrium point. 

\begin{figure}
\includegraphics{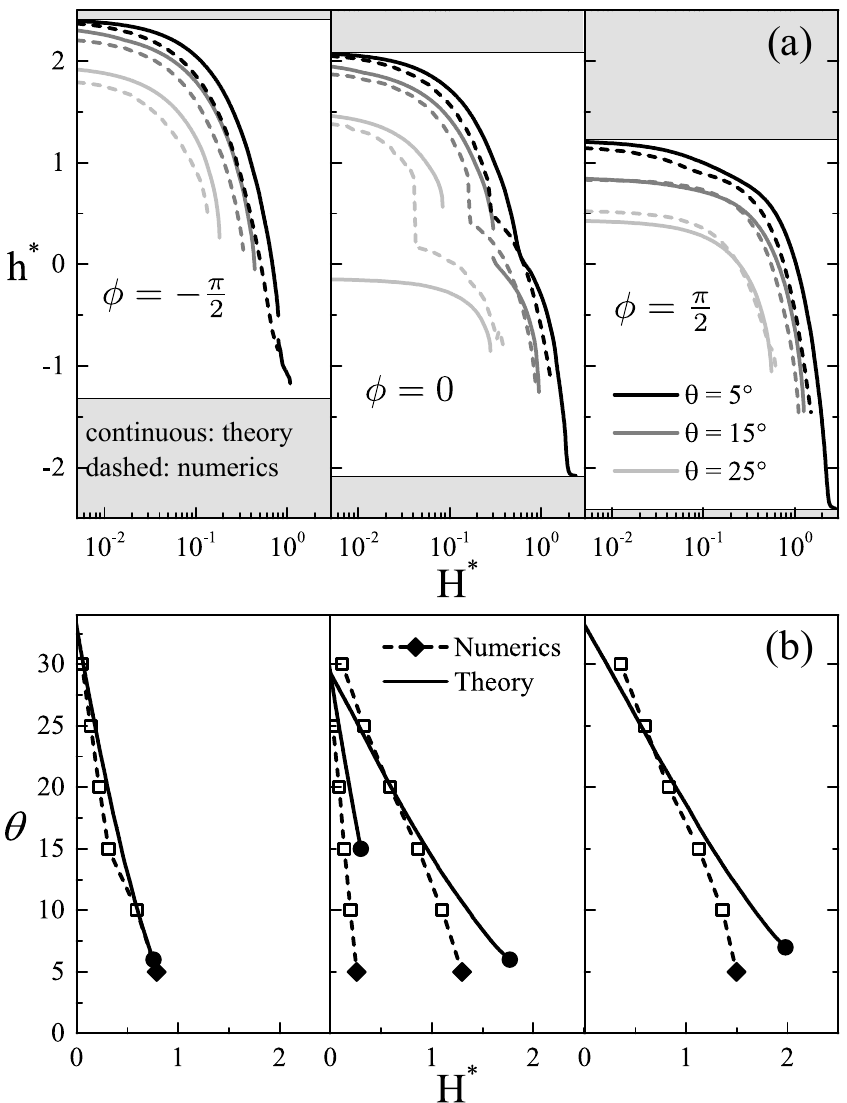}%
\caption{a) Comparison of the desorption curves obtained for the egg-carton pattern from the theory (continuous lines, 
stable solutions only) and energy minimization (dashed lines) for the three morphologies shown in Fig.~\ref{fig.stats_egg_carton}.
$h^*$ is the film level normalized w.r.t. the rms roughness, and $H^*$ is the film curvature normalized w.r.t. the rms curvature.
Grey areas represent values outside the upper ($h_+$) and lower ($h_-$) bounds of the patterns.
Results are shown for $\theta =$ $5^{\circ}$ (black), $15^{\circ}$ (dark grey), and $25^{\circ}$ (light grey).
b) Wetting phase diagrams (same parameters as in the three top panels). The transition lines from the theory 
and energy minimizations are shown as continuous lines and squares, respectively (dashed lines are guides to the eye).
The transition lines are obtained by locating the discontinuities in the desorption curves, and end at critical end points, as indicated by the closed symbols.}
\label{fig.desorption_egg_h}
\end{figure}

Fig.~\ref{fig.desorption_egg_h}a shows a series of adsorption isotherms, in terms of the average level of the  
contact line $h$ as function of the curvature $H$. They demonstrate satisfactory agreement between the interface 
displacement approach (solid curves) and the numerical energy minimizations (dashed curves) for all cases considered. 
For a saddle point located either close to the peaks or to the valleys of the topography  ($\phi=\pm \pi/2$),
the desorption curves exhibit a single discontinuous jump from a solution of finite film thickness
to the dry state. This transition reflects the fact that the topography cannot accommodate the previous  
liquid configuration anymore at the given $H$ and $\theta$ on the substrate topography.
The transition line is analogous to the off-coexistence filling transitions predicted
for simple translationally invariant substrates \cite{Rascon1999,Rodriguez-Rivas2014}. 
 
The largest allowed curvature corresponds to the curvature of the topography in the single periodic 
valley of the egg-carton, and is attained for vanishing $\theta$. 
In this case the adsorption isotherm exhibits a continuous behavior.
Note that in the case $\phi=-\pi/2$, the liquid
film consists of a single connected domain, while in the case $\phi=\pi/2$, the liquid film consists
of a periodic arrangement of identical isolated domains. As long as as the statistical properties are similar, our analysis
shows that for a periodic substrate the topology of the liquid domain plays no role in the adsorption
or desorption. The case $\phi=0$, where the saddle point is located around $z=0$, introduces
a relevant qualitative difference, as drying proceeds with a change of topology
when the liquid interface dewets the saddle point. Consequently, we observe
an additional discontinuity,  splitting the desorption curve into two branches, both characterized 
by a finite amount of adsorbed liquid. Note that this feature is captured in both the interface displacement approach and the numerical calculations.
    
It is worth noting that the interface displacement approach captures the presence of the two branches, because the statistical properties are strongly affected by the presence of the saddle point
(cf. Fig.~\ref{fig.stats_egg_carton}).
On the other hand, a close look at the numerically computed shape of the liquid interface shows 
that the contact line  experiences a stronger deviation in the vertical direction in the vicinity of the saddle point.
Such configuration departs from the approximation that length of the contact line is approximated by the length of surface isolines, as required by the interface displacement approach to obtain Eq.~(\ref{eq.approximation}). 
Consequently, the agreement is slightly less accurate. 

Fig.~\ref{fig.desorption_egg_h}b shows the resulting phase diagrams in the plane 
spanned by $\theta$ and $H$. The solid curves are obtained from the interface 
displacement model, while the squares re\-present the discontinuities in the curves 
displayed in the top panel. Obviously, the qualitative agreement is very good, as 
the number of observed discontinuities is the same in all three bottom panels.
The quantitative agreement concerning  the intercept and slope of  the phase 
boundaries is satisfactory as well, given the crudeness of the approximations used in the interface displacement model.
 
 So far we can conclude that the interface displacement model put forward earlier \cite{Herminghaus2012,Herminghaus2012a} is capable of reproducing the structure of the adsorption isotherms rather well, although substantial simplifications have been involved in the derivation of its central equation (\ref{eq.master_equation}).

\section{\label{sec:irregular}Irregular roughness}
 
Let us now turn to more irregular roughness topographies. For the numerical investigations, we have to ensure that the boundary conditions are met at the limits of the support. 
This can be best achieved by imposing periodic boundary conditions. 
Consequently,  we generated periodic roughness profiles $f(\mathbf{r})$
by means of a superposition of many ($> 400$) sinusoidal modes shifted by randomly chosen phases
$\phi_{\textbf{k}} \in \left\lbrace 0, 2\pi \right\rbrace$ 
\begin{equation}
\label{eq.fourier}
f(\textbf{r})=\sum_{\textbf{k}} a_{\textbf{k}} \sin \left( \textbf{k} \cdot \textbf{r} + \phi_{\textbf{k}}	\right)
\end{equation}
with $\textbf{r} \equiv (x,y) \in \Omega \equiv \left[0,1\right] \times \left[0,1\right]$ and two dimensional wave vectors
$\textbf{k} = 2\pi(m,n)$ where $m,n \in \left\lbrace -10,...,10 \right\rbrace$. 
The real Fourier amplitudes decay according to a power law,
$$ a_{\textbf{k}} =
\left\lbrace
\begin{array}{ll}
0  & \mathrm{for} \; \mathbf{k}=0\\
A_0 (1+|\mathbf{k}^{2}|)^{-1/2} & \mathrm{else}\\
\end{array}\right.$$
This leads to topographies with nearly Gaussian properties (see below).

In order to create topographies with non-Gaussian properties, we distort the so-obtained profile by applying the transformation
\begin{equation}
\label{eq.transform}
\tilde{f}(\textbf{r}) = f(\textbf{r}) + \epsilon \cdot f(\textbf{r})^2.
\end{equation}
The effect of Eq.~(\ref{eq.transform}) is to sharpen the peaks and flatten the valleys of $\tilde{f}$ as compared to $f$.
For $\epsilon=0$ the pattern is unchanged and has (nearly) Gaussian properties. 
For $\epsilon>0$ we depart from Gaussian properties as the height distribution becomes skewed, and a positive correlation is introduced between the elevation $z$ and surface slope.
A measure for the deviation of the chosen profile from Gaussian properties is the third moment (skewness) $Sk$ of the height distribution, which vanishes for a perfectly Gaussian topography \cite{Flannery2007, Polycarpou2004}.
We studied the cases $\epsilon = 0$, $\epsilon = 0.02$, and $\epsilon = 0.05$.
   
\begin{figure}[t,b]
\includegraphics{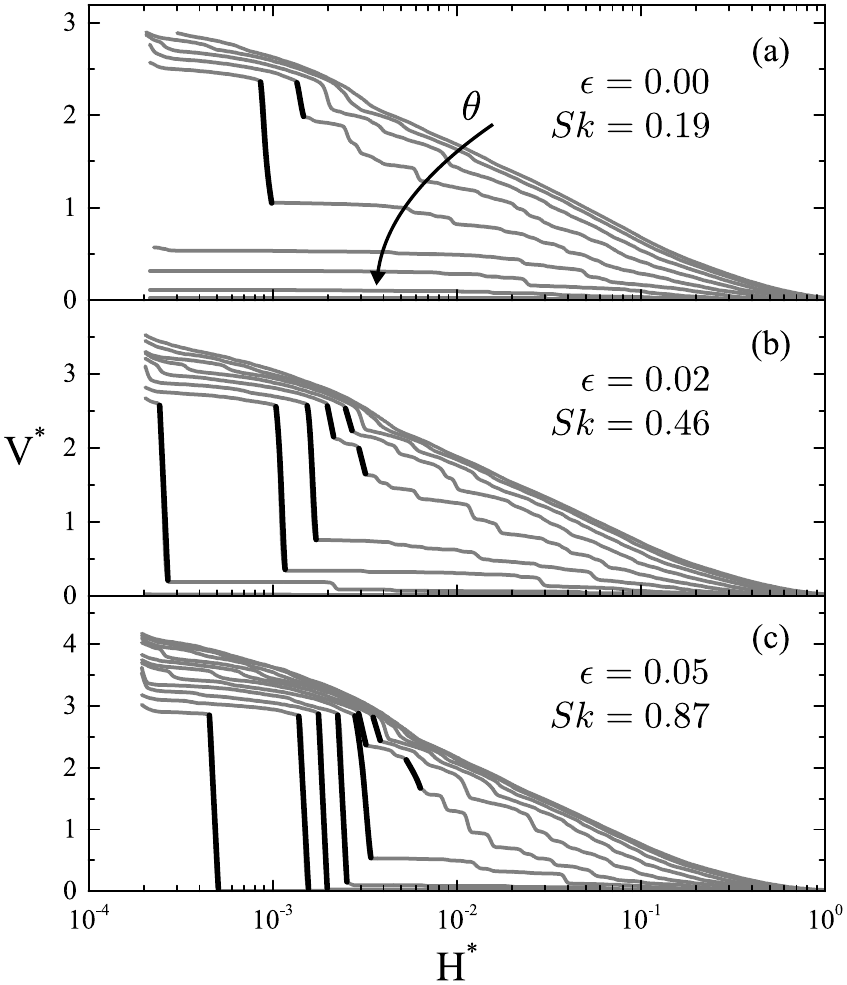}
\caption{\label{fig.desorption_random_vol}
Example of desorption curves obtained from 
the numerical energy minimization for $\epsilon = 0$ (a), $\epsilon = 0.02$ (b) and $\epsilon = 0.05$ (c). $V^*$ is the adsorbed volume per unit area, normalized with the rms roughness amplitude.
In each panel, the contact angle starts at $\theta 1^\circ$ and increases in one-degree steps. 
The black portions of the curves connect the desorption branches through the discontinuity (i.e. circles reported in Fig.~\ref{fig.morph_diag_random}).
}
\end{figure}
     
\begin{figure*}[t,b]
\includegraphics{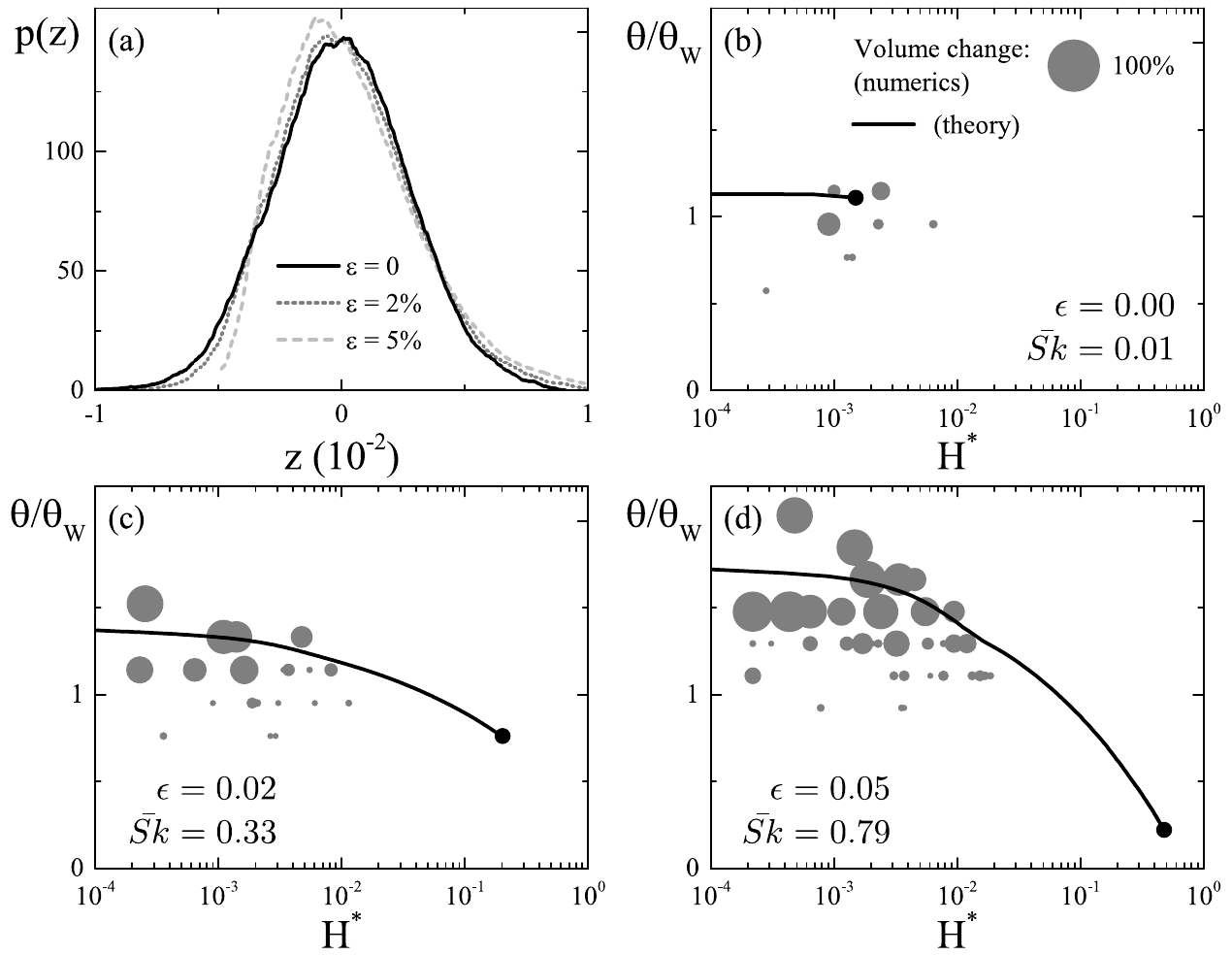}%
\caption{(a) Height distribution for the numerically generated random roughness with $\epsilon$ from 0 to 0.05 (cf. Eq.~(\ref{eq.transform})). 
(b-d) Comparison between the theory and numerical results. We plot the phase diagram spanned by the normalized curvature $H^*$ and the ratio $\theta/\theta_W$, with $\theta_W$ = 5.2$^{\circ}$, 5.3$^{\circ}$ and 5.4$^{\circ}$ for panels b-d respectively.
From the numerical results, the diameter of each circle is proportional to the height of the jump. 
The open circle, for scale, corresponds to a 100\% jump.
Note that only events assessing for at least 10\% of the total volume are shown.
The continuous black line is the transition predicted by the interface displacement model.
\label{fig.morph_diag_random} 
}
\end{figure*}

A typical set of results for the energy minimizations is reported in Fig.~\ref{fig.desorption_random_vol}.
In the case of a nearly Gaussian topography ($\epsilon=0$) we observe only a few discontinuities
for a finite curvature, and only for an intermediate range of material contact angles.
These jumps involve also a relatively small volume variation, indicating that after 
the transition many troughs remain filled with liquid, and continuously dry out until 
the largest curvature allowed is reached. This finding can be related to the presence of saddle points at any 
level of the topography, whose cumulative effect involves a dramatic change in the topology of the liquid interface 
from a single or a few large domains to a large number of small domains.
Both for larger and smaller contact angle
we observe a continuous drying, but while for small $\theta$ the drying starts from a completely
filled roughness, for large $\theta$ the initial state, for zero curvature, involves already a partially dry topography.
 
On the other hand, for the case $\epsilon=0.05$, we observe a large number of jumps at intermediate curvatures,
most of which lead directly to the dry solution. This finding shows that the stability of the upper branches 
of the liquid morphologies is reduced when most of the saddle points are located close to the bottom.

In contrast to the egg-carton topography, the location of discontinuities from numerical
energy minimizations on an irregular topography, as in the present case, can be affected 
by strong fluctuations related to local details. To improve  comparability, we therefore 
created ten sample topographies by assigning different series of random phases, and performed ensemble averaging.
The interface displacement model is then applied to the equivalent pattern obtained 
by combining the statistical properties of all ten samples. 
For $\epsilon = [0, 0.02, 0.05]$, the corresponding ensemble averages
for the skewness are $\bar{Sk}=[0.006, 0.330, 0.790]$.
The results from the energy minimizations instead are analysed by detecting 
the curvature $H$ and the magnitude $\Delta V$ of the volume difference. 
Each jump is then represented by a circle in Fig.~\ref{fig.morph_diag_random}, centered at the point where the jump was found, and with a diameter proportional to $\Delta V$.
Note that the contact angle is normalized by the respective Wenzel angle $\theta_W$ for each topography.

As we can observe in Fig.~\ref{fig.morph_diag_random}, the position of the transition line predicted by the interface displacement approach is consistent with the discontinuities identified with the
energy minimizations.
For the case $\epsilon=0$ we observe a short flat line, surrounded by a few spots of small radius.
Contrarily for the case $\epsilon=0.05$ we observe a clear transition line, surrounded by large spots.
Remarkably, Fig.~\ref{fig.morph_diag_random}b-d shows that 
in the limit $H^* \rightarrow 0$, the transition lines and numerical events depart from a contact angle $\theta_f$ larger than the Wenzel angle $\theta_W$.
This behavior is in agreement with the interface displacement theory \cite{Herminghaus2012a} (see also discussion below).
In particular, for the nearly Gaussian configuration 
($\epsilon = 0$), we expect $\theta_f/\theta_W \approx 2/\sqrt{\pi} \approx 1.13$, in very good agreement with Fig.~\ref{fig.morph_diag_random}b.

\section{\label{sec:random_roughness}Nearly Gaussian vs real roughness}

In the previous section, we validated the interface displacement model for a complex irregular roughness, but we were still limited to a relatively small number of corrugation modes.
In the present section, we apply the interface displacement approach to more complex substrates, 
in order to compare a nearly Gaussian roughness to the real topography of a sand blasted surface. 

\begin{figure*}
\includegraphics{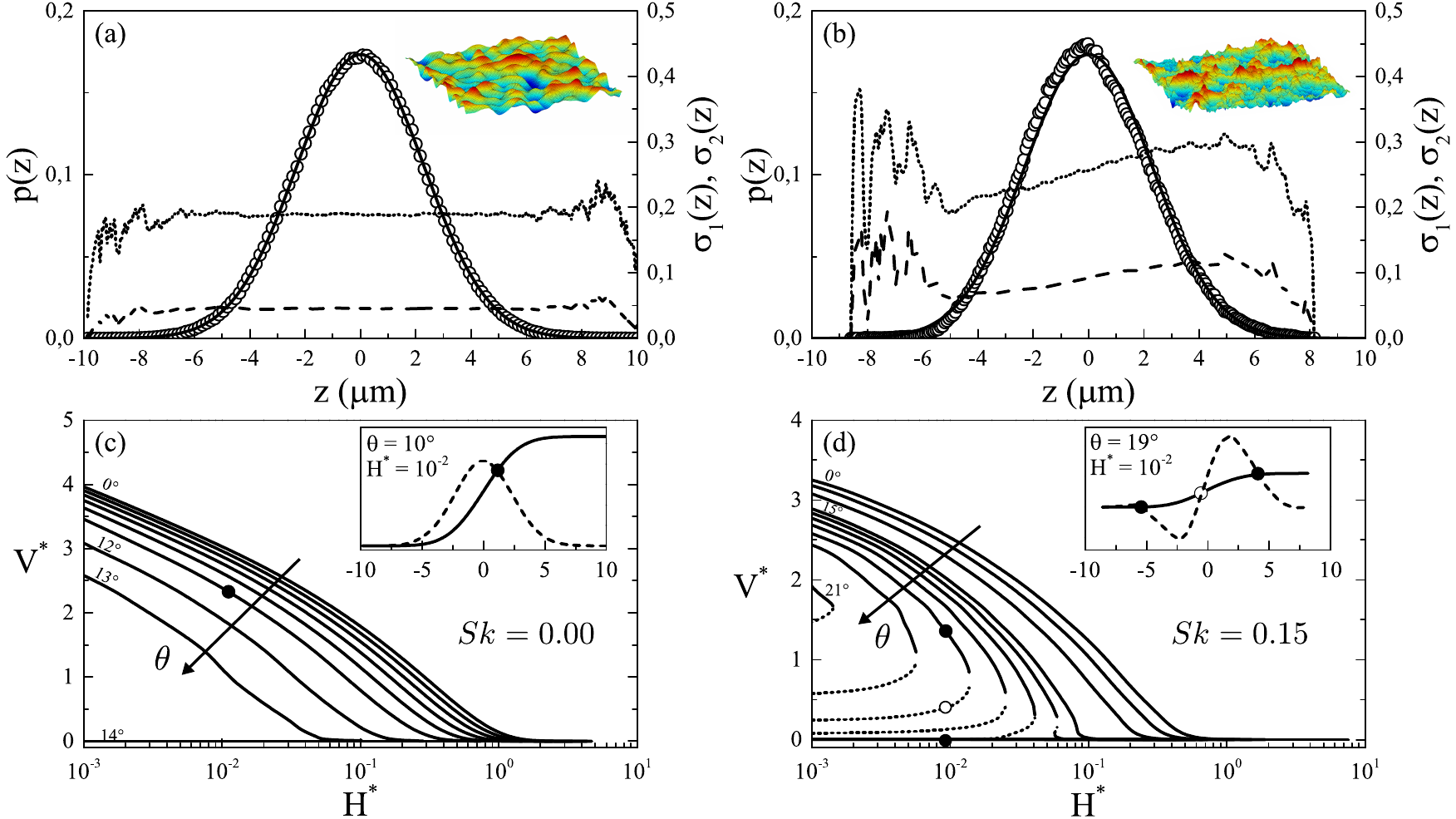}
\caption{Statistical properties of a random roughness with Gaussian power spectrum (a) and of a 
real rough sample treated by sandblasting (b).
Height distribution $p(z)$ are shown as open circles and overlays the analytical Gaussian curve in black. 
Conditional mean slope $\sigma_1(z)$ and $\sigma_2(z)$ are shown as dotted and dashed lines respectively.
Inset represents the unit periodic cell in (a) and a 400 x 500 $\mu m$ window in (b).
(c) and (d) show the corresponding desorption isotherms.
Continuous and dotted curves correspond to stable and unstable solutions to Eq.~(\ref{eq.master_equation}), 
respectively. Insets are example of graphical resolution with closed/open circles corresponding to 
stable/unstable solutions and mapped onto the corresponding curves.
}
\label{fig.stat_prop_and_desorption}
\end{figure*}

Let us first consider an ideal Gaussian roughness. 
We assume the profile to have vanishing mean, rms roughness $a$, and correlation length $\tau$.
The height distribution is then given by $p(z) = \mathcal{N}(0,a^2)$, where $\mathcal{N}(0,a^2)$ denotes the Gaussian normal distribution with variance $a^2$ and zero mean. Additionally, $\sigma_1 = \sqrt{\pi}a/\tau$ and $\sigma_2 = 4 (a/\tau)^2$ are constants \cite{Herminghaus2012, Herminghaus2012a}, indicating that the probability of finding 
a certain slope at a given level, $z$, is independent of $z$. This is linked to the fact that an ideal 
Gaussian roughness has an infinite support. Consequently, Eq.~(\ref{eq.master_equation}) admits a unique 
solution for $\theta < \theta_f = \frac{\sigma_2}{\sigma_1} = \frac{4 a}{\sqrt{\pi}\tau}$ ($\forall H$), 
and no solution for $\theta > \theta_f$. Therefore the phase diagram of a ideal Gaussian roughness 
is characterized by a horizontal line at $\theta = \theta_f$ delimiting the transition from wet to dry configurations.
It is instructive to compare $\theta_f$ with the Wenzel angle $\theta_W$. For Gaussian roughness, 
the Wenzel parameter is $r \approx 1+(a/\tau)^2$ \cite{Bhushan2008}. 
This leads to $\theta_W \approx 2a/\tau$ and to a fixed ratio 
$\theta_f/\theta_W \approx 2/\sqrt{\pi}$. 
For a non-Gaussian roughness, we also expect $\theta_f$ to be systematically 
larger than $\theta_W$ \cite{Herminghaus2012a}, as we already observed in the previous 
section  (cf. Fig.~\ref{fig.morph_diag_random}b-d). In practice, this means that if 
the liquid does not wet the substrate well enough to fulfill the Wenzel condition 
for complete wetting (Eq.~(\ref{eq.wenzel})), it may nevertheless form a wetting layer. 
 In the present case, 
we expect the transition from dry to wet being dominated by complex energy barriers 
as described for wicking and imbibition phenomena \cite{semprebon2014}.

 We investigate the numerical solution of Eq.~(\ref{eq.master_equation}) on nearly Gaussian roughness, 
by limiting the amplitude of the topography in the real space.
The surface is obtained by taking the inverse Fourier transform of an isotropic Gaussian power spectrum,
\begin{equation}
\label{eq.gaussian_spectrum}
S(\mathbf{k})=\frac{a^2 \tau^2}{2} |\mathbf{k}| \exp \left( \frac{-|\mathbf{k}|^2\tau^2}{4}   \right)
\end{equation}
where $\mathbf{k} = (k_x,k_y)$ is the 2D wave vector. The rms amplitude $a$ and 
correlation length $\tau$ are set to 2.3 $\mu$m and 21.5 $\mu$m, to match the properties 
of the topography of the the experimental sample. The latter is obtained by sandblasting 
of a copper plate, with sand grains of size in the range $70-110 \mu$m (Rohde AG).
The surface profile is measured with a white light interferometer (Weeko NT1100) 
over an area of 450*600 $\mu$m, with lateral and vertical resolutions of 
$\delta x \approx 0.8 \mu$m and $\delta z< 1$ nm respectively.
The map of the topography is pre-processed by applying a $5*5$ pixels Gaussian 
filter in order to remove measurement artifacts.
The statistical properties are computed from profiles obtained 
at three different positions on the sample.

Both the surface profiles and their statistical properties $p(z)$ and $\sigma_{1,2}(z)$ are shown in 
Fig.~\ref{fig.stat_prop_and_desorption}, panels a) and b). As expected for the Gaussian roughness,
the height distribution matches the Gaussian distribution, and $\sigma_{1,2}$ appear constant 
over most part of the range of $z$, in agreement with expected values of $\sigma_1=0.190$ and 
$\sigma_2=0.046$ respectively.
Because of the limited support, there is a highest elevation at $z = z_+$ and a deepest trough at $z=z_-$. As a consequence, both $\sigma_{1}$ and $\sigma_{2}$ drop to zero as $z \rightarrow z_{+/-}$. 
The height distribution of the real sample seems at first glance close to the 
analytical Gaussian curve. However, a difference is found in the value of the 
skewness, $Sk \approx 0.15$, indicating a predominance of valleys.
This departure from Gaussian properties is also reflected in the positive 
correlation between the slope and the elevation. Such feature can be 
expected if one considers the roughening mechanism underlying sandblasting, 
which is known to produce positively skewed topographies \cite{Whitehouse1974, Griffiths2003, Bellucci2012}.
The desorption curves obtained from the numerical solution of Eq.~(\ref{eq.master_equation}) 
are shown in Fig.~\ref{fig.stat_prop_and_desorption} panels c) and d). 
The adsorbed volume  per unit area is normalized by the rms amplitude $a$.

Two noticeable differences arise between a nearly Gaussian roughness on finite support 
and an ideal one. For a nearly Gaussian roughness we observe that the adsorbed volume 
at coexistence is finite $V^*(H=0) = V^*(z_+)$ and the topography completely dries for a finite value of $H^*$.
Instead, for an ideal Gaussian one we would have $V^*(H=0)\rightarrow\infty$, while the complete drying would occur 
for $H^*\rightarrow\infty$, because of the non vanishing probability of finding troughs of any size and curvature.
Regardless of these two aspects, the other features are present. Fig.~\ref{fig.stat_prop_and_desorption}c 
exhibits no bifurcation and the desorption curves continuously decrease until reaching the curvature of 
the smallest trough. These results, compared to the nearly Gaussian topography considered in the previous 
section (Fig.~\ref{fig.morph_diag_random}a), suggest that the few discontinuities observed were an effect 
of the limited number of modes considered
\footnote{For the patterns investigated in section \ref{sec:irregular}, statistics on the ensemble for 
$\epsilon=0$ agree with Gaussian properties. However because of the small sample size, individual samples 
can still have have non-negligible skewness and exhibit jumps in the desorption curves. This is for example 
the case in Fig.~\ref{fig.desorption_random_vol}a where for $\epsilon=0$ the sample skewness is 0.19}.
Increasing the contact angle, the largest value of $\theta$ 
for which Eq.~(\ref{eq.master_equation}) admits a solution is $13^{\circ}$.
This corresponds to the innermost curve in Fig.~\ref{fig.stat_prop_and_desorption}c.
If $\theta$ is slightly increased to just $14^{\circ}$, no adsorption is observed anymore.
Hence we can identify a critical value of $\theta$ above which the adsorbed amount of liquid 
goes abruptly to zero, irrespective of the mean curvature, $H$. 
In the case of Gaussian roughness this value has an analytical expression. Its value is 
$\theta_f = \frac{\sigma_2}{\sigma_1}=\frac{4a}{\sqrt{\pi} \tau}=13.8^{\circ}$.

\begin{figure}[t,b]
\includegraphics{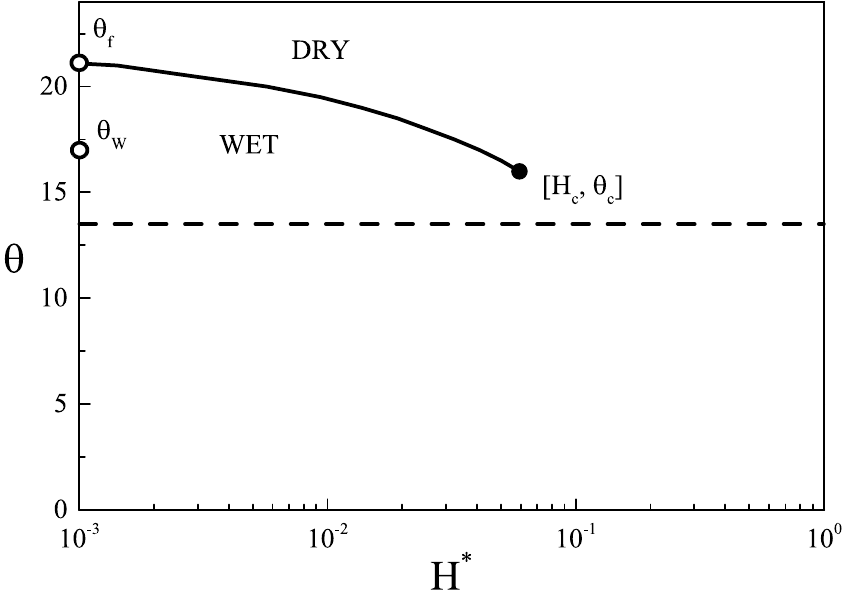}
\caption {Phase diagram for the cases displayed in Fig.~\ref{fig.stat_prop_and_desorption}.
The horizontal dashed line indicates the critical angle $\theta_c = 13.5(\pm 0.5)^{\circ}$ which can be deduced from  for the Gaussian pattern (Fig.~\ref{fig.stat_prop_and_desorption}a and \ref{fig.stat_prop_and_desorption}c).
The continuous black line describes the discontinuous transition as introduced in Ref. \cite{Herminghaus2012}, 
obtained for the sandblasted surface ((Fig.~\ref{fig.stat_prop_and_desorption}b and \ref{fig.stat_prop_and_desorption}d)). 
Crossing this line leads to a discontinuous jump in the amount of adsorbed liquid.}
\label{fig.morph_diag_real}
\end{figure}

The predictions of the interface displacement model applied to the sandblasted topography are shown in Fig.~\ref{fig.stat_prop_and_desorption}d.
Consistently with the measured statistical properties of the topography, the desorption curves behave in agreement with the findings presented in sec. \ref{sec:random_roughness}, for the case $\epsilon=0.05$. 
For small values of $\theta$, the model predicts a continuous decrease of $V$ with $H^*$. 
As $\theta$ increases, approaching the critical value $\theta_c\approx 15^\circ$, a bifurcation point emerges at $H_c^* \approx 0.06$.
For $H^* > H_c^*$ no solution is detected, while for $H^*<H_c^*$ both stable and unstable branches of Eq.~(\ref{eq.master_equation}) appear.
The largest material angle admitting a stable solution in the case of the real roughness is $\theta_f \approx 21^\circ$ (inner left curve in Fig.~\ref{fig.stat_prop_and_desorption}d).
Finally, we present in Fig.~\ref{fig.morph_diag_real} the phase diagram obtained for the cases displayed in Fig.~\ref{fig.stat_prop_and_desorption}, summarizing all features described above.
Remarkably, for the real roughness we find that the critical end point $[H^*_c,\theta_c]$ lies in the physical region of the parameter space.

\section{\label{sec:conclusion}Conclusions}

In this paper we reported on numerical investigations of wetting on various roughness topographies.
Our results compare favourably with predictions made earlier a purely analytic interface displacement approach. 
Specifically, we corroborated that filling transitions are generic on random roughness topographies, and the phase diagrams are in reasonable quantitative agreement with the predictions of the analytic theory.
Furthermore, our results support the startling prediction that Gaussian random topographies behave qualitatively differently.
First, we find no filling transitions in adsorption isotherms on perfectly Gaussian roughness.  Second, there is a filling transition when $\theta$ is varied, but its position is independent of $H$. 
This confirms that modeling topographic roughness with a Gaussian random process may miss important physical aspects of wetting.

\begin{acknowledgments}
The authors acknowledge generous support from
BP International Inc. within the ExploRe research program,
and funding from DFG within the Grant no. HE 2016/14-2 of SPP 1486 'PiKo'.
We thank Martin Brinkmann for inspiring discussions.
\end{acknowledgments}

\bibliography{biblio}

\begin{thebibliography}{27}%
\makeatletter
\providecommand \@ifxundefined [1]{%
 \@ifx{#1\undefined}
}%
\providecommand \@ifnum [1]{%
 \ifnum #1\expandafter \@firstoftwo
 \else \expandafter \@secondoftwo
 \fi
}%
\providecommand \@ifx [1]{%
 \ifx #1\expandafter \@firstoftwo
 \else \expandafter \@secondoftwo
 \fi
}%
\providecommand \natexlab [1]{#1}%
\providecommand \enquote  [1]{``#1''}%
\providecommand \bibnamefont  [1]{#1}%
\providecommand \bibfnamefont [1]{#1}%
\providecommand \citenamefont [1]{#1}%
\providecommand \href@noop [0]{\@secondoftwo}%
\providecommand \href [0]{\begingroup \@sanitize@url \@href}%
\providecommand \@href[1]{\@@startlink{#1}\@@href}%
\providecommand \@@href[1]{\endgroup#1\@@endlink}%
\providecommand \@sanitize@url [0]{\catcode `\\12\catcode `\$12\catcode
  `\&12\catcode `\#12\catcode `\^12\catcode `\_12\catcode `\%12\relax}%
\providecommand \@@startlink[1]{}%
\providecommand \@@endlink[0]{}%
\providecommand \url  [0]{\begingroup\@sanitize@url \@url }%
\providecommand \@url [1]{\endgroup\@href {#1}{\urlprefix }}%
\providecommand \urlprefix  [0]{URL }%
\providecommand \Eprint [0]{\href }%
\providecommand \doibase [0]{http://dx.doi.org/}%
\providecommand \selectlanguage [0]{\@gobble}%
\providecommand \bibinfo  [0]{\@secondoftwo}%
\providecommand \bibfield  [0]{\@secondoftwo}%
\providecommand \translation [1]{[#1]}%
\providecommand \BibitemOpen [0]{}%
\providecommand \bibitemStop [0]{}%
\providecommand \bibitemNoStop [0]{.\EOS\space}%
\providecommand \EOS [0]{\spacefactor3000\relax}%
\providecommand \BibitemShut  [1]{\csname bibitem#1\endcsname}%
\let\auto@bib@innerbib\@empty
\bibitem [{\citenamefont {Herminghaus}(2012{\natexlab{a}})}]{Herminghaus2012}%
  \BibitemOpen
  \bibfield  {author} {\bibinfo {author} {\bibfnamefont {S.}~\bibnamefont
  {Herminghaus}},\ }\href {\doibase 10.1140/epje/i2012-12043-8} {\bibfield
  {journal} {\bibinfo  {journal} {Eur. Phys. J. E}\ }\textbf {\bibinfo {volume}
  {35}},\ \bibinfo {pages} {43} (\bibinfo {year}
  {2012}{\natexlab{a}})}\BibitemShut {NoStop}%
\bibitem [{\citenamefont {Herminghaus}(2012{\natexlab{b}})}]{Herminghaus2012a}%
  \BibitemOpen
  \bibfield  {author} {\bibinfo {author} {\bibfnamefont {S.}~\bibnamefont
  {Herminghaus}},\ }\href {\doibase 10.1103/PhysRevLett.109.236102} {\bibfield
  {journal} {\bibinfo  {journal} {Phys. Rev. Lett.}\ }\textbf {\bibinfo
  {volume} {109}},\ \bibinfo {pages} {236102} (\bibinfo {year}
  {2012}{\natexlab{b}})}\BibitemShut {NoStop}%
\bibitem [{\citenamefont {Seemann}\ \emph {et~al.}(2005)\citenamefont
  {Seemann}, \citenamefont {Brinkmann}, \citenamefont {Kramer}, \citenamefont
  {Lange},\ and\ \citenamefont {Lipowsky}}]{Seemann2005}%
  \BibitemOpen
  \bibfield  {author} {\bibinfo {author} {\bibfnamefont {R.}~\bibnamefont
  {Seemann}}, \bibinfo {author} {\bibfnamefont {M.}~\bibnamefont {Brinkmann}},
  \bibinfo {author} {\bibfnamefont {E.~J.}\ \bibnamefont {Kramer}}, \bibinfo
  {author} {\bibfnamefont {F.~F.}\ \bibnamefont {Lange}}, \ and\ \bibinfo
  {author} {\bibfnamefont {R.}~\bibnamefont {Lipowsky}},\ }\href {\doibase
  10.1073/pnas.0407721102} {\bibfield  {journal} {\bibinfo  {journal} {Proc.
  Nat. Acad. Sci. USA}\ }\textbf {\bibinfo {volume} {102}},\ \bibinfo {pages}
  {1848} (\bibinfo {year} {2005})}\BibitemShut {NoStop}%
\bibitem [{\citenamefont {Rejmer}\ \emph {et~al.}(1999)\citenamefont {Rejmer},
  \citenamefont {Dietrich},\ and\ \citenamefont
  {Napi\'{o}rkowski}}]{Rejmer1999}%
  \BibitemOpen
  \bibfield  {author} {\bibinfo {author} {\bibfnamefont {K.}~\bibnamefont
  {Rejmer}}, \bibinfo {author} {\bibfnamefont {S.}~\bibnamefont {Dietrich}}, \
  and\ \bibinfo {author} {\bibfnamefont {M.}~\bibnamefont {Napi\'{o}rkowski}},\
  }\href {\doibase 10.1103/PhysRevE.60.4027} {\bibfield  {journal} {\bibinfo
  {journal} {Phys. Rev. E}\ }\textbf {\bibinfo {volume} {60}},\ \bibinfo
  {pages} {4027} (\bibinfo {year} {1999})}\BibitemShut {NoStop}%
\bibitem [{\citenamefont {Rasc\'{o}n}\ and\ \citenamefont
  {Parry}(2000)}]{rascon2000}%
  \BibitemOpen
  \bibfield  {author} {\bibinfo {author} {\bibfnamefont {C.}~\bibnamefont
  {Rasc\'{o}n}}\ and\ \bibinfo {author} {\bibfnamefont {A.~O.}\ \bibnamefont
  {Parry}},\ }\href {\doibase 10.1038/35039590} {\bibfield  {journal} {\bibinfo
   {journal} {Nature}\ }\textbf {\bibinfo {volume} {407}},\ \bibinfo {pages}
  {986} (\bibinfo {year} {2000})}\BibitemShut {NoStop}%
\bibitem [{\citenamefont {Malijevsk\'{y}}\ and\ \citenamefont
  {Parry}(2013)}]{parry2013}%
  \BibitemOpen
  \bibfield  {author} {\bibinfo {author} {\bibfnamefont {A.}~\bibnamefont
  {Malijevsk\'{y}}}\ and\ \bibinfo {author} {\bibfnamefont {A.~O.}\
  \bibnamefont {Parry}},\ }\href {\doibase 10.1103/PhysRevLett.110.166101}
  {\bibfield  {journal} {\bibinfo  {journal} {Phys. Rev. Lett.}\ }\textbf
  {\bibinfo {volume} {110}},\ \bibinfo {pages} {166101} (\bibinfo {year}
  {2013})}\BibitemShut {NoStop}%
\bibitem [{\citenamefont {Parry}\ \emph {et~al.}(2014)\citenamefont {Parry},
  \citenamefont {Malijevsk\'{y}},\ and\ \citenamefont
  {Rasc\'{o}n}}]{parry2014}%
  \BibitemOpen
  \bibfield  {author} {\bibinfo {author} {\bibfnamefont {A.~O.}\ \bibnamefont
  {Parry}}, \bibinfo {author} {\bibfnamefont {A.}~\bibnamefont
  {Malijevsk\'{y}}}, \ and\ \bibinfo {author} {\bibfnamefont {C.}~\bibnamefont
  {Rasc\'{o}n}},\ }\href {\doibase 10.1103/PhysRevLett.113.146101} {\bibfield
  {journal} {\bibinfo  {journal} {Phys. Rev. Lett.}\ }\textbf {\bibinfo
  {volume} {113}},\ \bibinfo {pages} {146101} (\bibinfo {year}
  {2014})}\BibitemShut {NoStop}%
\bibitem [{\citenamefont {Malijevsk\'{y}}\ and\ \citenamefont
  {Parry}(2015)}]{parry2015}%
  \BibitemOpen
  \bibfield  {author} {\bibinfo {author} {\bibfnamefont {A.}~\bibnamefont
  {Malijevsk\'{y}}}\ and\ \bibinfo {author} {\bibfnamefont {A.~O.}\
  \bibnamefont {Parry}},\ }\href {\doibase 10.1103/PhysRevE.91.052401}
  {\bibfield  {journal} {\bibinfo  {journal} {Phys. Rev. E}\ }\textbf {\bibinfo
  {volume} {91}},\ \bibinfo {pages} {052401} (\bibinfo {year}
  {2015})}\BibitemShut {NoStop}%
\bibitem [{\citenamefont {Persson}\ \emph {et~al.}(2005)\citenamefont
  {Persson}, \citenamefont {Albohr}, \citenamefont {Tartaglino}, \citenamefont
  {Volokitin},\ and\ \citenamefont {Tosatti}}]{Persson2005}%
  \BibitemOpen
  \bibfield  {author} {\bibinfo {author} {\bibfnamefont {B.~N.~J.}\
  \bibnamefont {Persson}}, \bibinfo {author} {\bibfnamefont {O.}~\bibnamefont
  {Albohr}}, \bibinfo {author} {\bibfnamefont {U.}~\bibnamefont {Tartaglino}},
  \bibinfo {author} {\bibfnamefont {A.~I.}\ \bibnamefont {Volokitin}}, \ and\
  \bibinfo {author} {\bibfnamefont {E.}~\bibnamefont {Tosatti}},\ }\href
  {\doibase 10.1088/0953-8984/17/1/R01} {\bibfield  {journal} {\bibinfo
  {journal} {J. Phys. Condens. Matter}\ }\textbf {\bibinfo {volume} {17}},\
  \bibinfo {pages} {R1} (\bibinfo {year} {2005})}\BibitemShut {NoStop}%
\bibitem [{\citenamefont {Savva}\ \emph {et~al.}(2010)\citenamefont {Savva},
  \citenamefont {Kalliadasis},\ and\ \citenamefont
  {Pavliotis}}]{Grigorios2010}%
  \BibitemOpen
  \bibfield  {author} {\bibinfo {author} {\bibfnamefont {N.}~\bibnamefont
  {Savva}}, \bibinfo {author} {\bibfnamefont {S.}~\bibnamefont {Kalliadasis}},
  \ and\ \bibinfo {author} {\bibfnamefont {G.~A.}\ \bibnamefont {Pavliotis}},\
  }\href {\doibase 10.1103/PhysRevLett.104.084501} {\bibfield  {journal}
  {\bibinfo  {journal} {Phys. Rev. Lett.}\ }\textbf {\bibinfo {volume} {104}},\
  \bibinfo {pages} {084501} (\bibinfo {year} {2010})}\BibitemShut {NoStop}%
\bibitem [{\citenamefont {Chakraborty}\ \emph {et~al.}(2012)\citenamefont
  {Chakraborty}, \citenamefont {Dingari},\ and\ \citenamefont
  {Chakraborty}}]{Chakraborty2012}%
  \BibitemOpen
  \bibfield  {author} {\bibinfo {author} {\bibfnamefont {D.}~\bibnamefont
  {Chakraborty}}, \bibinfo {author} {\bibfnamefont {N.~N.}\ \bibnamefont
  {Dingari}}, \ and\ \bibinfo {author} {\bibfnamefont {S.}~\bibnamefont
  {Chakraborty}},\ }\href {\doibase 10.1021/la303603c} {\bibfield  {journal}
  {\bibinfo  {journal} {Langmuir}\ }\textbf {\bibinfo {volume} {28}},\ \bibinfo
  {pages} {16701} (\bibinfo {year} {2012})}\BibitemShut {NoStop}%
\bibitem [{\citenamefont {David}\ and\ \citenamefont
  {Neumann}(2013)}]{Wilhelm2013}%
  \BibitemOpen
  \bibfield  {author} {\bibinfo {author} {\bibfnamefont {R.}~\bibnamefont
  {David}}\ and\ \bibinfo {author} {\bibfnamefont {A.~W.}\ \bibnamefont
  {Neumann}},\ }\href {\doibase 10.1021/la400294t} {\bibfield  {journal}
  {\bibinfo  {journal} {Langmuir}\ }\textbf {\bibinfo {volume} {29}},\ \bibinfo
  {pages} {4551} (\bibinfo {year} {2013})}\BibitemShut {NoStop}%
\bibitem [{\citenamefont {Bottiglione}\ and\ \citenamefont
  {Carbone}(2015)}]{Carbone2015}%
  \BibitemOpen
  \bibfield  {author} {\bibinfo {author} {\bibfnamefont {F.}~\bibnamefont
  {Bottiglione}}\ and\ \bibinfo {author} {\bibfnamefont {G.}~\bibnamefont
  {Carbone}},\ }\href {\doibase 10.1088/0953-8984/27/1/015009} {\bibfield
  {journal} {\bibinfo  {journal} {J. Phys. Condens. Matter}\ }\textbf {\bibinfo
  {volume} {27}},\ \bibinfo {pages} {015009} (\bibinfo {year}
  {2015})}\BibitemShut {NoStop}%
\bibitem [{\citenamefont {Wenzel}(1936)}]{Wenzel1936}%
  \BibitemOpen
  \bibfield  {author} {\bibinfo {author} {\bibfnamefont {R.~N.}\ \bibnamefont
  {Wenzel}},\ }\href {\doibase 10.1021/ie50320a024} {\bibfield  {journal}
  {\bibinfo  {journal} {Ind. Eng. Chem.}\ }\textbf {\bibinfo {volume} {28}},\
  \bibinfo {pages} {988} (\bibinfo {year} {1936})}\BibitemShut {NoStop}%
\bibitem [{\citenamefont {Longuet-Higgins}(1957)}]{Higgins1957}%
  \BibitemOpen
  \bibfield  {author} {\bibinfo {author} {\bibfnamefont {M.~S.}\ \bibnamefont
  {Longuet-Higgins}},\ }\href {\doibase 10.1098/rsta.1957.0018} {\bibfield
  {journal} {\bibinfo  {journal} {Phys. Trans. R. Soc. A}\ }\textbf {\bibinfo
  {volume} {250}},\ \bibinfo {pages} {157} (\bibinfo {year}
  {1957})}\BibitemShut {NoStop}%
\bibitem [{\citenamefont {Brakke}(1996)}]{Brakke1996}%
  \BibitemOpen
  \bibfield  {author} {\bibinfo {author} {\bibfnamefont {K.~A.}\ \bibnamefont
  {Brakke}},\ }\href {\doibase 10.1098/rsta.1996.0095} {\bibfield  {journal}
  {\bibinfo  {journal} {Phil. Trans. R. Soc. A}\ }\textbf {\bibinfo {volume}
  {354}},\ \bibinfo {pages} {2143} (\bibinfo {year} {1996})}\BibitemShut
  {NoStop}%
\bibitem [{\citenamefont {Semprebon}\ \emph {et~al.}(2012)\citenamefont
  {Semprebon}, \citenamefont {Herminghaus},\ and\ \citenamefont
  {Brinkmann}}]{Semprebon2012}%
  \BibitemOpen
  \bibfield  {author} {\bibinfo {author} {\bibfnamefont {C.}~\bibnamefont
  {Semprebon}}, \bibinfo {author} {\bibfnamefont {S.}~\bibnamefont
  {Herminghaus}}, \ and\ \bibinfo {author} {\bibfnamefont {M.}~\bibnamefont
  {Brinkmann}},\ }\href {\doibase 10.1039/c2sm25156f} {\bibfield  {journal}
  {\bibinfo  {journal} {Soft Matter}\ }\textbf {\bibinfo {volume} {8}},\
  \bibinfo {pages} {6301} (\bibinfo {year} {2012})}\BibitemShut {NoStop}%
\bibitem [{\citenamefont {Rasc{\'{o}}n}\ \emph {et~al.}(1999)\citenamefont
  {Rasc{\'{o}}n}, \citenamefont {Parry},\ and\ \citenamefont
  {Sartori}}]{Rascon1999}%
  \BibitemOpen
  \bibfield  {author} {\bibinfo {author} {\bibfnamefont {C.}~\bibnamefont
  {Rasc{\'{o}}n}}, \bibinfo {author} {\bibfnamefont {a.~O.}\ \bibnamefont
  {Parry}}, \ and\ \bibinfo {author} {\bibfnamefont {a.}~\bibnamefont
  {Sartori}},\ }\href {\doibase 10.1103/PhysRevE.59.5697} {\bibfield  {journal}
  {\bibinfo  {journal} {Phys. Rev. E}\ }\textbf {\bibinfo {volume} {59}},\
  \bibinfo {pages} {5697} (\bibinfo {year} {1999})},\ \Eprint
  {http://arxiv.org/abs/9902070} {9902070} \BibitemShut {NoStop}%
\bibitem [{\citenamefont {Rodr{\'{\i}}guez-Rivas}\ \emph
  {et~al.}(2014)\citenamefont {Rodr{\'{\i}}guez-Rivas}, \citenamefont
  {Galv{\'{a}}n},\ and\ \citenamefont {Romero-Enrique}}]{Rodriguez-Rivas2014}%
  \BibitemOpen
  \bibfield  {author} {\bibinfo {author} {\bibfnamefont {A.}~\bibnamefont
  {Rodr{\'{\i}}guez-Rivas}}, \bibinfo {author} {\bibfnamefont {J.}~\bibnamefont
  {Galv{\'{a}}n}}, \ and\ \bibinfo {author} {\bibfnamefont {J.~M.}\
  \bibnamefont {Romero-Enrique}},\ }\href {\doibase
  10.1088/0953-8984/27/3/035101} {\bibfield  {journal} {\bibinfo  {journal} {J.
  Phys. Condens. Matter}\ }\textbf {\bibinfo {volume} {27}},\ \bibinfo {pages}
  {035101} (\bibinfo {year} {2014})},\ \Eprint {http://arxiv.org/abs/1408.1013}
  {1408.1013} \BibitemShut {NoStop}%
\bibitem [{\citenamefont {Press}\ \emph {et~al.}(2007)\citenamefont {Press},
  \citenamefont {Teukolsly}, \citenamefont {Vetterling},\ and\ \citenamefont
  {Flannery}}]{Flannery2007}%
  \BibitemOpen
  \bibfield  {author} {\bibinfo {author} {\bibfnamefont {W.~H.}\ \bibnamefont
  {Press}}, \bibinfo {author} {\bibfnamefont {S.~A.}\ \bibnamefont
  {Teukolsly}}, \bibinfo {author} {\bibfnamefont {W.~T.}\ \bibnamefont
  {Vetterling}}, \ and\ \bibinfo {author} {\bibfnamefont {B.~P.}\ \bibnamefont
  {Flannery}},\ }\href@noop {} {\emph {\bibinfo {title} {{Numerical Recipes:
  The Art of Scientific Computing}}}},\ \bibinfo {edition} {3rd}\ ed.\
  (\bibinfo  {publisher} {Cambridge University Press},\ \bibinfo {year}
  {2007})\ Chap.~\bibinfo {chapter} {14}, p.\ \bibinfo {pages}
  {994}\BibitemShut {NoStop}%
\bibitem [{\citenamefont {Tayebi}\ and\ \citenamefont
  {Polycarpou}(2004)}]{Polycarpou2004}%
  \BibitemOpen
  \bibfield  {author} {\bibinfo {author} {\bibfnamefont {N.}~\bibnamefont
  {Tayebi}}\ and\ \bibinfo {author} {\bibfnamefont {A.~A.}\ \bibnamefont
  {Polycarpou}},\ }\href {\doibase 10.1016/j.triboint.2003.11.010} {\bibfield
  {journal} {\bibinfo  {journal} {Tribol. Int.}\ }\textbf {\bibinfo {volume}
  {37}},\ \bibinfo {pages} {491} (\bibinfo {year} {2004})}\BibitemShut
  {NoStop}%
\bibitem [{\citenamefont {Nosonovsky}\ and\ \citenamefont
  {Bhushan}(2008)}]{Bhushan2008}%
  \BibitemOpen
  \bibfield  {author} {\bibinfo {author} {\bibfnamefont {M.}~\bibnamefont
  {Nosonovsky}}\ and\ \bibinfo {author} {\bibfnamefont {B.}~\bibnamefont
  {Bhushan}},\ }\href {\doibase 10.1088/0953-8984/20/22/225009} {\bibfield
  {journal} {\bibinfo  {journal} {J. Phys. Condens. Matter}\ }\textbf {\bibinfo
  {volume} {20}},\ \bibinfo {pages} {225009} (\bibinfo {year}
  {2008})}\BibitemShut {NoStop}%
\bibitem [{\citenamefont {Semprebon}\ \emph {et~al.}(2014)\citenamefont
  {Semprebon}, \citenamefont {Forsberg}, \citenamefont {Priest},\ and\
  \citenamefont {Brinkmann}}]{semprebon2014}%
  \BibitemOpen
  \bibfield  {author} {\bibinfo {author} {\bibfnamefont {C.}~\bibnamefont
  {Semprebon}}, \bibinfo {author} {\bibfnamefont {P.}~\bibnamefont {Forsberg}},
  \bibinfo {author} {\bibfnamefont {C.}~\bibnamefont {Priest}}, \ and\ \bibinfo
  {author} {\bibfnamefont {M.}~\bibnamefont {Brinkmann}},\ }\href {\doibase
  10.1039/c4sm00684d} {\bibfield  {journal} {\bibinfo  {journal} {Soft Matter}\
  }\textbf {\bibinfo {volume} {10}},\ \bibinfo {pages} {5739} (\bibinfo {year}
  {2014})}\BibitemShut {NoStop}%
\bibitem [{\citenamefont {Whitehouse}(1974)}]{Whitehouse1974}%
  \BibitemOpen
  \bibfield  {author} {\bibinfo {author} {\bibfnamefont {D.~J.}\ \bibnamefont
  {Whitehouse}},\ }\href {\doibase 10.1007/978-1-4613-4490-2} {\emph {\bibinfo
  {title} {{Characterization of Solid Surfaces. Chapter 3}}}},\ edited by\
  \bibinfo {editor} {\bibfnamefont {P.~F.}\ \bibnamefont {Kane}}\ and\ \bibinfo
  {editor} {\bibfnamefont {G.~B.}\ \bibnamefont {Larrabee}}\ (\bibinfo
  {publisher} {Springer US},\ \bibinfo {address} {Boston, MA},\ \bibinfo {year}
  {1974})\BibitemShut {NoStop}%
\bibitem [{\citenamefont {Griffiths}(2003)}]{Griffiths2003}%
  \BibitemOpen
  \bibfield  {author} {\bibinfo {author} {\bibfnamefont {B.}~\bibnamefont
  {Griffiths}},\ }\href {\doibase 10.1016/B978-185718033-6/50020-3} {\emph
  {\bibinfo {title} {{Engineering Drawing for Manufacture}}}}\ (\bibinfo
  {publisher} {Elsevier},\ \bibinfo {year} {2003})\ pp.\ \bibinfo {pages}
  {111--133}\BibitemShut {NoStop}%
\bibitem [{\citenamefont {Monetta}\ and\ \citenamefont
  {Bellucci}(2012)}]{Bellucci2012}%
  \BibitemOpen
  \bibfield  {author} {\bibinfo {author} {\bibfnamefont {T.}~\bibnamefont
  {Monetta}}\ and\ \bibinfo {author} {\bibfnamefont {F.}~\bibnamefont
  {Bellucci}},\ }\href {\doibase 10.4236/ojrm.2012.13007} {\bibfield  {journal}
  {\bibinfo  {journal} {OJRM}\ }\textbf {\bibinfo {volume} {01}},\ \bibinfo
  {pages} {41} (\bibinfo {year} {2012})}\BibitemShut {NoStop}%
\bibitem [{Note1()}]{Note1}%
  \BibitemOpen
  \bibinfo {note} {For the patterns investigated in section \ref
  {sec:irregular}, statistics on the ensemble for $\epsilon =0$ agree with
  Gaussian properties. However because of the small sample size, individual
  samples can still have have non-negligible skewness and exhibit jumps in the
  desorption curves. This is for example the case in Fig.~\ref
  {fig.desorption_random_vol}a where for $\epsilon =0$ the sample skewness is
  0.19}\BibitemShut {NoStop}%
\end{thebibliography}%

\end{document}